\documentclass[pre,twocolumn,aps,showpacs]{revtex4-1}

\usepackage{amsmath}
\usepackage{amssymb}
\usepackage{graphicx}
\usepackage{color}

\newcommand{\ri}{R_I}
\newcommand{\eps}{\epsilon}
\newcommand{\gat}{\gamma_a}
\newcommand{\bv}{{\mathbf v}}
\newcommand{\bV}{{\mathbf V}}

\newcommand{\hatn}{\hat{n}}

\begin{document}

\title{Non-equilibrium fluctuations in frictional granular motor: experiments
and kinetic theory}

\author{Andrea Gnoli}
\email{andrea.gnoli@isc.cnr.it}
\affiliation{Istituto dei Sistemi Complessi - CNR and Dipartimento di Fisica,
Universit\`a ''Sapienza'', p.le A. Moro 2, 00185 Rome, Italy}
\affiliation{Istituto dei Sistemi Complessi - CNR, via del Fosso del Cavaliere
100, 00133 Rome, Italy}

\author{Alessandro Sarracino}
\email{alessandro.sarracino@roma1.infn.it}
\affiliation{Istituto dei Sistemi Complessi - CNR and Dipartimento di Fisica,
Universit\`a ''Sapienza'', p.le A. Moro 2, 00185 Rome, Italy}

\author{Alberto Petri}
\email{alberto.petri@isc.cnr.it}
\affiliation{Istituto dei Sistemi Complessi - CNR, via del Fosso del Cavaliere
100, 00133 Rome, Italy}

\author{Andrea Puglisi}
\email{andrea.puglisi@roma1.infn.it}
\affiliation{Istituto dei Sistemi Complessi - CNR and Dipartimento di Fisica,
Universit\`a ''Sapienza'', p.le A. Moro 2, 00185 Rome, Italy}

\begin{abstract}
We report the study of a new experimental granular Brownian motor,
inspired to the one published in [Phys. Rev. Lett. 104, 248001
  (2010)], but different in some ingredients. As in that previous
work, the motor is constituted by a rotating pawl whose surfaces
break the rotation-inversion symmetry through alternated patches of
different inelasticity, immersed in a gas of granular
  particles. The main novelty of our  experimental
  setup is in the orientation of the main axis, which is parallel to
the (vertical) direction of shaking of the granular fluid,
guaranteeing an isotropic distribution for the velocities of colliding
grains, characterized by a variance $v_0^2$. We also keep the granular system
diluted, in order to
compare with Boltzmann-equation-based kinetic theory.  In agreement
with theory, we observe for the first time the crucial role of Coulomb friction
which
induces two main regimes: (i) rare collisions (RC), with an average
drift $\langle \omega \rangle \sim v_0^3$, and (ii) frequent
collisions (FC), with $\langle \omega \rangle \sim v_0$. We also study
the fluctuations of the angle spanned in a large time interval,
$\Delta \theta$, which in the FC regime is proportional to the work
done upon the motor. We observe that the Fluctuation Relation is
satisfied with a slope which weakly depends on the relative collision
frequency.
\end{abstract}


\maketitle

\section{Introduction}

Brownian motors (BM) are devices that can rectify thermal
fluctuations, in order to perform work against external
loads~\cite{R02,HM03}. The basic underlying mechanism relies on the
presence of non-equilibrium conditions, breaking the time-reversal
symmetry in the dynamics, together with some spatial anisotropy, which
allows unidirectional motion. Although these general constraints are
clearly understood~\cite{f63}, many open questions remain to be
answered, concerning the several different mechanisms for the
realization of such devices. For instance, given a particular shape of
the probe, the prediction of the drift
direction is far from obvious, in
particular if several sources of dissipation are present in the
system, inducing competitive effects.

Recent years have seen an increasing wide interest
on Brownian motors directly inspired to the original setup of the
Feynman's ratchet~\cite{f63}.  In these ``collisional Brownian
motors'' (CBM) fluctuations are induced by the collisions of an
asymmetric probe with particles of molecular fluids at different
temperatures~\cite{BKM04,MBG04,BMK05}.  Since the presence of
dissipation is a fundamental ingredient to induce non-equilibrium
conditions, a natural framework where these kinds of systems have been
studied is the realm of granular media, where interactions do not
conserve energy due to inelasticity. Several
experimental~\cite{FTVV99,EWLM10,BDLPP11,HMPTGJ12} and theoretical
results~\cite{CB07,CPB07,CE08,CPB09} have been obtained for these
systems.

More recently, another source of dissipation has been shown to play an
important role in the dynamics of CBM: the Coulomb (or dry)
friction~\cite{TWV11,TBV11,GPDGPSP13}. Its main effect is the
introduction of two dynamical regimes, where the behavior of the
systems is dominated by collisions or friction, respectively.  More
surprisingly, it has also been shown that the Coulomb friction itself
can be sufficient to drive a motor effect, even if the probe is in
contact with a single molecular fluid at equilibrium~\cite{GPDGPSP13,SGP13}.
The role of friction has also been studied in other systems showing
motor effects~\cite{GC10,BS12}, where fluctuations are introduced by
noise terms, which are not related to particle collisions.

The context of granular systems also paves the way to the realization
of experiments aimed at validating some important general relations
derived for nonequilibrium systems, such as the Fluctuation
Relation~\cite{BPRV08} or the Hatano-Sasa relation~\cite{HS01}. In
particular, in granular systems, where noise and time-scale separation
are often not fully under control and where some coarse-grained on the
accessible quantities is present, the experimental study of these
relations is very useful to assess such results in more general
situations~\cite{JLM12,N12,MN12}.

In this paper we consider a new experimental setup for a frictional
granular CBM, in order to get closer to conditions where kinetic
theory can be applied. Moreover, at variance with previous studies, we
also take into account the presence of Coulomb friction which induces
interesting behaviors. This allows us to compare experimental results
with analytical predictions of kinetic theory. Furthermore, we focus
on the study of the non-equilibrium fluctuations of the spanned angle
in a time interval, which is related to the work done by the CBM. Our
findings suggest that a symmetry relation for these fluctuations is
verified in our system, in agreement with previous results for similar
experiments~\cite{JLM12}.

\section{Setup}

\begin{figure}[htbp]
\begin{center}
\includegraphics[angle=0,width=8cm,clip=true]{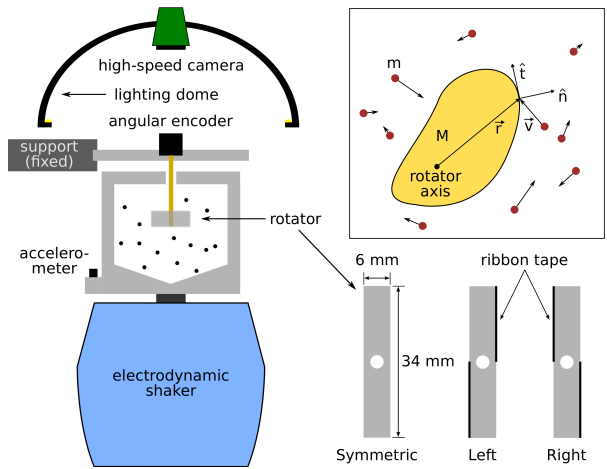}
\caption{{\bf Color online}: the experimental setup
\label{fig:setup}}
\end{center}
\end{figure}

The two main components of our setup are the granular gas and the
rotor, see Fig.~\ref{fig:setup} for visual explanation.  The
granular gas is made of $N=50$ spheres of polyoxymethylene (diameter
$d=6$ mm and mass $m=0.15$ g) contained in a
polymethyl-methacrylate (PMMA) cylinder of circular base, with area $A
\approx 6.36\times 10^3$ mm$^2$ and maximum height $70$ mm. The
cylinder is shaken with a sinusoidal signal at $53$ Hz and
variable amplitude, which is measured by the maximum
acceleration $\Gamma=a_{max}/G$ rescaled by the gravity acceleration  $G$.  The
velocity distribution of the spheres on the plane perpendicular to the
rotation axis is obtained by particle tracking via a fast camera
(see~\cite{PGGSV12} for details on the procedure) and is fairly
approximated by a Gaussian,
\begin{equation}
p_g(v) \sim e^{-v_x^2/(2v_0^2)},
\end{equation} 
where the ``thermal'' velocity $v_0$ has been introduced and $v_x$ may
be replaced by $v_y$ because the system is isotropic on the $\hat{xy}$
plane. Small deviations from the Gaussian are observed but are
neglected for the purpose of this study; see~\cite{GPDGPSP13} for
details.  We have changed $\Gamma$ from $5$ to $21$, finding for $v_0$
values from $120$ mm$^2$ s$^{-2}$ to
$500$ mm$^2$ s$^{-2}$.  The vertical average profile of the
gas density is close to the so-called ``leading order''
distribution~\cite{k98}: $n_0(z)=\frac{NG}{A
  v_0^2}\exp\left(\frac{-G}{v_0^2}z\right)$, where $z$ is the
coordinate of the vertical axis. This is sufficient to evaluate the
average density surrounding the rotator, estimated to be $n=n_0(z^*)$
where $z^*$ is the mid-heigth of the rotator.

Suspended into the gas, we have a pawl (also called ``rotator'')
rotating around a vertical axis. The rotator is a PMMA cylinder with a
rectangular base (see lower-right inset of Fig.~\ref{fig:setup}) of
perimeter $S$, height $h$ and total surface of the sides $\Sigma=S h
=1.2\times 10^{-3}$ mm$^2$. The rotator (including its axis) has mass
$M=6.49$ g and moment of inertia
$I=353$ g mm$^2$. The axis of the rotator is suspended to two
spheres bearing. The position vs time of the rotator is recorded by an
angular encoder (Avago Technologies, model AEDA-330), enclosing the
bearings. In Fig.~\ref{fig:setup}, top-right inset, the definition of
some quantities useful for the theory can be found. The study of the
Brownian motor phenomenon is obtained by applying insulating tape to
the rotator, partially covering its two largest surfaces, see
lower-right inset of Fig.~\ref{fig:setup} for an explanation. Turning
upside-down the rotator allows one to invert its chirality, defining two
possible orientations which we call ``right'' and ``left''.

A close analysis of the dynamics of the rotator shows that the angular velocity
$\omega$ is well
described by the following equation of motion:
\begin{equation} \label{lang1}
\dot{\omega}(t)= -\Delta \sigma[\omega(t)] -\gamma_a \omega(t) + \eta_{coll}(t)
\end{equation}
where $\Delta=F_{frict}/{I}=38\pm 4$ s$^{-2}$ is the
frictional force rescaled by inertia,
$\gamma_a=6\pm 1$ s$^{-1}$ is some viscous damping rate
related perhaps to air or to other dissipations in the bearings, and
$\eta_{coll}(t)$ is the random force due to collisions with the
granular gas particles. The granular gas itself is stationary and
(roughly) homogeneous with a mean free path proportional to
$\lambda=(n\Sigma)^{-1}$.  The pawl is further characterized by its
symmetric shape factor $\langle g^2\rangle_{surf}=1.51$
(see~\cite{CE08} for details), with $\langle \rangle_{surf}$ being a
uniform average over the surface of the object parallel to the
rotation axis and $g(s)=\frac{{\mathbf r}(s)\cdot \hat{t}(s)}{\ri}$
with $\hat{t}(s)=\hat{z} \times \hat{n}(s)$ which is the unit vector
tangent to the surface at the point ${\mathbf r}(s)$, and $\hat{n}(s)$
is the unit vector perpendicular to the surface at that point. We have
also introduced the radius of inertia $\ri=\sqrt{I/M}$ of the
rotator. We refer to the top-right inset of Fig.~\ref{fig:setup} for a
visual explanation of symbols. The restitution coefficient between
spheres and the pawl has been measured to be
$\alpha_+ \approx 0.67$ on the PMMA naked
surface and $\alpha_- \approx 0.35$ on the
tape-covered face.

It is useful to introduce the ``equipartition'' angular velocity
$\omega_0 = v_0 \eps/\ri$ where $\eps=\sqrt{\frac{m}{M}}$. Note that,
because of inelastic collisions and frictional dissipations, the
rotator {\em does not} satisfy equipartition and $\omega_0$ is only a
useful reference value.  It is natural to adimensionalize the rotator
angular velocity defining $\Omega=\frac{\omega}{\omega_0}$.

\section{Theory}

The single particle probability density function (pdf) $p(\omega,t)$
for the angular velocity of the rotator is fully described, under the
assumption of diluteness which guarantees Molecular Chaos, by the
following linear Boltzmann equation~\cite{CE08,TWV11,GPDGPSP13}
\begin{subequations} \label{beq}
\begin{align} 
\partial_t p(\omega,t)&=\partial_\omega [(\Delta \sigma(\omega)+\gat
\omega)p(\omega,t)]+J[p,p_g] \\
J[p,p_g]&=\int d\omega' W(\omega|\omega')p(\omega',t)- p(\omega,t) f_c(\omega),
\\
f_c(\omega)&=\int d\omega' W(\omega'|\omega)\\
W(\omega'|\omega)&=\rho S \int \frac{ds}{S} \int d\bv p_g(\bv) \Theta[(\bV(s)-
\bv) \cdot \hatn] \times \\ \nonumber &|(\bV(s)- \bv) \cdot
\hatn|\delta[\omega'-\omega-\Delta
\omega(s)],\\
\Delta \omega(s) &= (1+\alpha)\frac{[\bV(s)- \bv] \cdot \hatn}{\ri}\frac{
g(s)\eps^2}{1+\eps^2g(s)^2}, \label{eq:colrule}
\end{align}
\end{subequations}
where we introduce the rates $W(\omega'|\omega)$ for the transition
$\omega \to \omega'$, the velocity-dependent collision frequency
$f_c(\omega)$, the pdf for the gas particle velocities $p_g(\bv)$ and
the so-called kinematic constraint in the form of Heaviside step
function $\Theta[(\bV- \bv) \cdot \hatn]$ which enforces the kinematic
condition necessary for impact.  Here ${\mathbf V}(s)=\omega
  \hat{z} \times {\mathbf r}(s)$ is the linear velocity of the rotator
  at the point of impact ${\mathbf r}(s)$.
The collision rule is implemented by
Eq.~\eqref{eq:colrule}~\cite{CE08}.

An estimate of the ratio between the stopping time due to dissipation
(dominated by dry friction) $\tau_\Delta \sim \frac{\omega_0}{
  \Delta}$ and the collisional time $\tau_c \sim \frac{1}{n \Sigma
  v_0}$ is given by the parameter
\begin{equation}
\beta^{-1} =\frac{\eps n\Sigma v_0^2}{\sqrt{2} \pi \ri
  \Delta} \approx\frac{\tau_\Delta}{\tau_c}.
\end{equation}
This parameter controls
the transition from a regime (at $\beta^{-1} \ll 1$) with fast
stopping due to dissipation, called RC (rare collisions), and a regime
(at $\beta^{-1} \gg 1$) with the rotator always in motion,
continuously perturbed by collisions, called FC (frequent collisions).


When the mass of the rotator is large with respect to the mass of the
granular gas particles, collisions are small perturbations to
$\omega(t)$ (see Eq.~\eqref{eq:colrule}). Then, it makes sense to
expand Eq.~\eqref{beq} in powers of $\eps$~\cite{CPB07,CE08}: by retaining only up to the
second derivative, a Fokker-Planck equation is obtained. The basic
result of this procedure is that the collisional noise in
Eq.~\eqref{lang1} is cast into the sum of a white noise $\eta(t)$ plus a viscous
drag and a systematic force inducing the motor effect:
\begin{equation} \label{etamap}
\eta_{coll}(t) \to \eta(t)-\gamma_g \omega(t)+\tau_{motor},
\end{equation}
with $\langle \eta \rangle =0$ and $\langle
\eta(t)\eta(t')\rangle=\Gamma_g \delta(t-t')$.  The expression
for $\gamma_g$, $\tau_{motor}$ and the amplitude of the noise $\Gamma_g$
have been obtained, for a generic asymmetric rotator in a dilute granular
gas, in ~\cite{CE08}. For our particular shape they read~\cite{TWV11}
\begin{align}
\gamma_g &=
\sqrt{\frac{2}{\pi}}\lambda^{-1}\eps^2 v_0\langle (1+\alpha) g^2
\rangle_{surf},\\
\tau_{motor}&= \gamma_g \sqrt{\frac{3 \pi}{32}} \frac{1}{\eps}
\frac{\alpha_+-\alpha_-}{2+\alpha_++\alpha_-},\\
\Gamma_g &= (1+\alpha)\gamma_g \frac{\eps^2}{R_I^2} v_0^2,
\end{align}
with $\alpha=(\alpha_++\alpha_-)/2$. We want to highlight, here, a
fundamental point concerning the collisional noise $\eta_{coll}$: as
Eq.~\eqref{etamap} explicitly shows, such a noise is in general not
white, and, even more importantly, {\em it is not independent} from
the instantaneous velocity $\omega$. This makes sense, as it is the
superposition of the variations of angular velocity due to collisions,
which, as shown by Eq.~\eqref{eq:colrule}, depends on $\omega$. For
this reason our model, described alternatively by Eq.~\eqref{lang1} or
Eq.~\eqref{beq}, is very different from a model -- apparently similar
-- recently introduced in~\cite{BS12}, as well as from other previous
models~\cite{BBG06}.

In the FC limit, $\beta^{-1} \gg 1$, the
Coulomb friction term and the external viscosity may be neglected, i.e.
\begin{equation}
\gamma_a \omega +\Delta \sigma(\omega) \ll \gamma_g \omega,
\end{equation}
so that Eq.~\eqref{lang1} is cast into the much simpler form
\begin{equation} \label{lang2}
\dot{\omega}(t)=-\gamma_g \omega(t) + \eta(t) + \tau_{motor}.
\end{equation}
From such an equation one may estimate the average velocity of the
Brownian motor to be~\cite{TWV11,TBV11}
\begin{equation} \label{drift_fc}
\langle \omega \rangle = \frac{\tau_{motor}}{\gamma_g}.
\end{equation}
Another interesting observation follows from
Eq.~\eqref{lang2}. It concerns the fluctuations $f(\Delta \theta)$ of
the spanned angle in a time interval of length $\Delta t$, $\Delta \theta =
\theta(t+\Delta t)-\theta(t)$ for any $t$ in the steady state. For the
particularly simple linear Langevin case, Eq.~\eqref{lang2}, it can be
shown that such fluctuations obey, for large $\Delta t$, the following
Fluctuation Relation (FR)~\cite{LS99}:
\begin{equation}
\phi(\Delta \theta) = \log\left[\frac{f(\Delta \theta)}{f(-\Delta
\theta)}\right] \approx s\Delta \theta,
\end{equation}
with
\begin{equation} \label{slope_th}
s=\frac{\gamma_g \tau_{motor}}{\Gamma_g} \approx \frac{\tau_{motor}}{\langle
\omega^2 \rangle}.
\end{equation}
We mention that such an FR is closely related to the FR for the
entropy produced in the time $\Delta t$, which in this system is
approximated by the work done by the ``Brownian motor force'' $W \approx
\tau_{motor} \Delta \theta$ divided by the ``temperature'' $\langle
\omega^2 \rangle$~\cite{seifert05}.

In the RC regime, on the other side, one may assume that the probe's dynamics is
a sequence of
independent kicks received at zero velocity, resulting in the
following formula for the adimensional average angular velocity~\cite{TWV11}:
\begin{align} \label{drift_rc}
\langle \Omega \rangle &= q \beta^{-1} \\
q &= \frac{\sqrt{\pi}}{4}\left[(1+\alpha_+)^2-(1+\alpha_-)^2
  \right] \\ &\times
\left(\frac{\tan^{-1}\sqrt{\xi}}{\sqrt{\xi}}-\frac{1}{1+\xi}\right),
\end{align}
where $\xi=m L^2/(4 I)$ and $L$ is the lenght of the pawl. We
note that for the {\em dimensional}
angular velocity this means $|\omega| \sim v_0^3 \eps$. In the RC
regime the behavior of $\phi(\Delta \theta)$ is unknown in
principle. A FR for the entropy production certainly exists, but we
are not aware of a simple relation between $\Delta \theta$ and the
entropy produced in a given time interval.

\section{Motor effect}

\begin{figure}[htbp]
\begin{center}
\includegraphics[angle=0,width=8cm,clip=true]{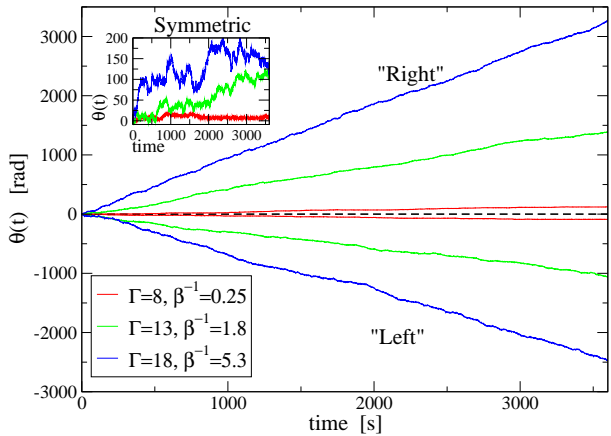}
\caption{{\bf Color online}: trajectories with different shaking parameters and
different chiralities. The inset shows the behavior of the
symmetric (i.e., without tape) pawl.
\label{fig:traj}}
\end{center}
\end{figure}

In Figure~\ref{fig:traj} the evolution in time of the angle
$\theta(t)$ spanned by the rotator is shown for different choices of
the maximum shaking acceleration $\Gamma$ and different orientation of
the asymmetry (``left'', L, or ``right'', R). A measure of the
``thermal'' velocity of the particles, $v_0$, through the fast camera,
allows one to determine $\beta^{-1}$ which estimates the relative
relevance of collisions with respect to Coulomb friction. A steady
drift, signaling the presence of Brownian motor effect, is observed
both in the friction dominated regime ($\beta^{-1}<1$) and in the
collisions dominated regime ($\beta^{-1}>1$). Turning the rotator
upside down, i.e. changing its asymmetry from $L$ to $R$ or viceversa,
inverts the sign of the drift: such an observation confirms that the
observed drift is caused by the asymmetry of $\alpha$, as expected
from the given theoretical arguments. A further confirmation that the
observed drift is due to the surface heterogeneity comes from the
study of the ``symmetric'' rotator, which has no patches of
insulating tape: on the same timescale  only a very weak drift is observed, much
weaker than the one observed with the L- and R-type rotators. We
impute such a weak bias measured with the symmetric rotator to
imperfections of the experimental setup: any kind of asymmetry
(e.g. not sufficiently precise vertical alignments, not perfectly
circular profile of the container, etc.) may induce {\em secondary}
motor effects.

\begin{figure}[htbp]
\begin{center}
\includegraphics[angle=0,width=8cm,clip=true]{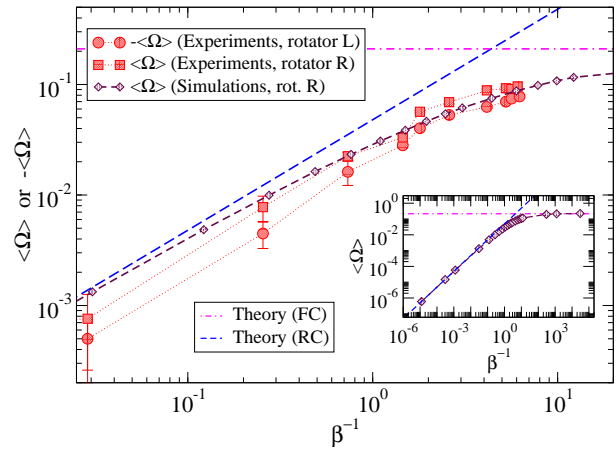}
\caption{{\bf Color online}: Average angular velocity, rescaled by
  $\omega_0$, of the rotator: experiments, theory and simulations, for
  several values of $\beta^{-1}$ and both chiralities. In the inset
  only simulations are show, in order to appreciate the comparison
  with both RC and FC theoretical limits at very small and very large
  $\beta^{-1}$.
\label{fig:ave}}
\end{center}
\end{figure}

In Figure~\ref{fig:ave}, the rescaled average angular velocity of the
rotator, $\langle \Omega \rangle$, measured in experiments with both
L- and R-type rotators, is shown as a function of $\beta^{-1}$. The
velocity of the L-rotator is changed of sign for the purpose of a
better visualization. In the same plot we have also shown the results
of numerical simulations of Eq.~\eqref{beq} with the same parameters
as in the experiment (only for the R-type rotator). Details about the
simulation can be found in the Supplemental Material of
ref.~\cite{GPDGPSP13}. As a first observation, we highlight the quite
good agreement between experimental and numerical data.

The thick blue dashed and magenta dot-dashed lines represent the theoretical
predictions for the RC and FC limits, Eqs.~\eqref{drift_rc}
and~\eqref{drift_fc} respectively.  In the inset of the Figure we have
displayed the results of the simulations on a much wider range of
$\beta^{-1}$, in order to appreciate the agreement with the
theoretical limits. The simulations teach us that such theoretical
predictions for the RC limit (FC limit) are useful for quite small
(large) values of $\beta^{-1}$. The experimentally accessible values
of $\beta^{-1}$ appear to be at the crossover between the two regimes:
nevertheless they span a sufficiently wide range, so that both the
$\langle \Omega \rangle \sim \beta^{-1}$ behavior (RC regime) and the
trend toward saturation $\langle \omega \rangle \to \textrm{const.}$
(FC regime) can be identified. We consider this to be the best
comparison, up to our present knowledge, between experimental granular
Brownian motors and kinetic theory. We mention that it is quite
difficult to expand the accessible $\beta^{-1}$ range. Indeed exploring
smaller values of $\beta^{-1}$ requires a considerable increase of the
dry friction coefficient $\Delta$, which is not under our direct
control; moreover, a large $\Delta$ may amplify non-ideal effects
where the behavior of the spheres bearings do not follow the Coulomb
law: such effects are already observed here (at small velocities small
deviations are observed) and are likely responsible for the not
perfect match with simulation results at small $\beta^{-1}$; we also
notice that {\em static friction} is not considered here, but it could
become relevant at large $\Delta$~\cite{TV11}. The opposite limit,
i.e. large values of $\beta^{-1}$, are even more difficult to be
attained, since they would require a larger collision frequency:
however the maximum acceleration of the shaker is a hard limit, while
increasing the number of grains does not trivially produces the
desired result, for two reasons: 1) higher densities correspond to a
{\em reduction} of the average kinetic energy and 2) Molecular Chaos
is only guaranteed at low density. Notwithstanding those limits, we
believe that the results of Fig.~\ref{fig:ave} are already a quite
good success of kinetic theory and make us claim that Eq.~\eqref{beq}
is a fair description of the experimental setup.

\section{Non-equilibrium fluctuations}

\begin{figure}[htbp]
\begin{center}
\includegraphics[angle=0,width=8cm,clip=true]{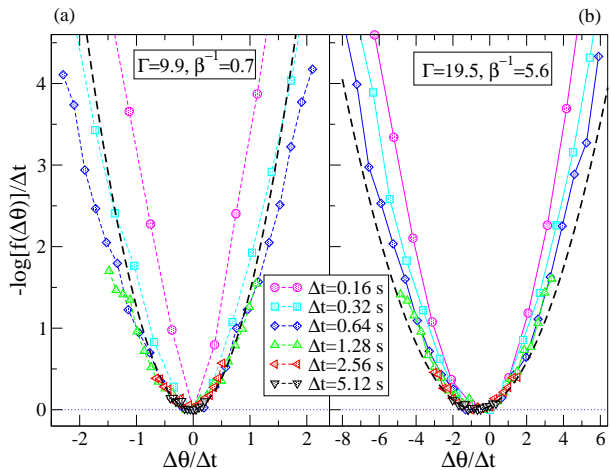}
\caption{{\bf Color online}: Empirical large deviation function for
  the fluctuations of $\Delta \theta$ for two experiments in the RC
  regime (left) and in the FC regime (right), at several values of
  $\Delta t$. A small correction is operated to each pdf $f(\Delta
  \theta)$, normalizing it by its maximum value, before taking the
  logarithm, for the purpose of a better vertical alignment.
\label{fig:ld}}
\end{center}
\end{figure}

In Fig.~\ref{fig:ld} we have displayed the empirical large deviation
rate (ldr) $-\log[f(\Delta \theta)]/\Delta t$ of the pdf $f(\Delta
\theta)$ for different choices of the time window $\Delta t$, in two
experiments with a small and a large value of $\beta^{-1}$. In both
cases we have also superimposed a parabolic fit of the data at the
largest available time. We mention that the characteristic time
$\tau_c$ for the decay of the angular velocity autocorrelation, not
shown here, is in the range $0.03-0.06$ s. As
frequently observed~\cite{PVTW06,PRV06}, the empirical large deviation
rate tends to become independent from time only at very large $\Delta
t \gg \tau_c$. Here we evaluate $\Delta t \sim 1$ s as a
minimum value before considering reached the large deviation limit. In
both experiments we may appreciate deviations from the parabolic fit,
i.e. slightly non-Gaussian tails, signaling that we are indeed probing
the large deviations of the pdf of $\Delta \theta$.

\begin{figure}[htbp]
\begin{center}
\includegraphics[angle=0,width=8cm,clip=true]{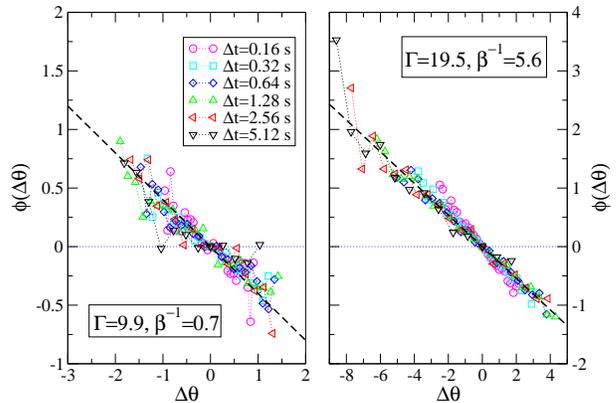}
\caption{{\bf Color online}: Asymmetry function $\phi(\Delta \theta)$ for the pdfs of $\Delta \theta$
 for two experiments in the RC regime (left) and in the FC regime (right), at several values of $\Delta t$.
\label{fig:gc}}
\end{center}
\end{figure}

\begin{figure}[htbp]
\begin{center}
\includegraphics[angle=0,width=8cm,clip=true]{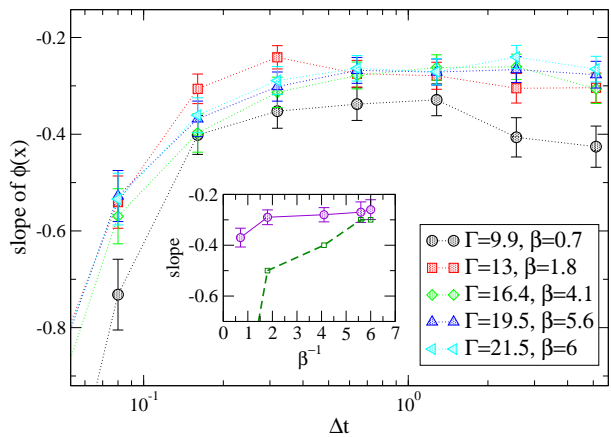}
\caption{{\bf Color online}: Slopes of the asymmetry functions for
  several different experiments, as a function of $\Delta t$. The
  inset shows the value of the slope (the values of each experiments
  are averaged over the plateau visible in the main plot at large
  $\Delta t$) as a function of $\beta^{-1}$. The squares joined by the
  green dashed line represent the result of formula~\eqref{slope_th}.
\label{fig:slopes}}
\end{center}
\end{figure}

The asymmetry function $\phi(\Delta \theta) = \log\left[\frac{f(\Delta
    \theta)}{f(-\Delta \theta)}\right]$ for the pdf of $\Delta \theta$
is shown, for the same two experiments, in Fig.~\ref{fig:gc}. At
values of $\Delta t$ large enough, but smaller than those necessary to
achieve a stable large deviation rate, the asymmetry functions already
display a linear behavior $\sim s \Delta \theta$ with a slope $s$ not
dramatically changing with time. The values of the slope have been
measured for all experiments and many values of $\Delta t$ and are
reported in Fig.~\ref{fig:slopes}.

While confirming the direct observation done in Fig.~\ref{fig:gc},
Fig.~\ref{fig:slopes} clearly teaches us that the slope of the
$\phi(\Delta \theta)$ becomes stable at times of the order $\Delta t
\sim 0.3$ s. In the inset of Fig.~\ref{fig:slopes} we have
displayed the slope $s$ (from a constant fit of the main plot
including values $\Delta t \ge 0.32$ s) vs. $\beta^{-1}$. The
squared symbols joined by the green dashed line represent the result
of formula~\eqref{slope_th}. Many important comments are in order
here: 1) for large values of $\beta^{-1}$, as expected, dry friction
becomes negligible and - at the leading order - the system is
described by Eq.~\eqref{lang2}, which is confirmed by the good
agreement of the slope with Eq.~\eqref{slope_th}; 2) at moderate and
small values of $\beta^{-1}$ the ``simplified'' Langevin description
of Eq.~\eqref{lang2} is not expected to hold, and indeed discrepancy
is found between experimental slopes of $\phi(\Delta \theta)$ and
those predicted by Eq.~\eqref{slope_th}; nevertheless 3) such
experimental values of the slope appear to depend only {\em weakly}
upon $\beta^{-1}$, so that they do not differ too much from the
values at large $\beta^{-1}$. The last observation is an empirical
fact which has not a simple explanation: at small values of
$\beta^{-1}$ the noise felt by the rotator is discontinuous and the
average drift can hardly be described as the effect of a continuous
torque (as it is $\tau_{motor}$ in the FC limit). Therefore it is not
clear at all how to define a work or an injected power and,
consequently, a candidate for the entropy production. A theory for the
fluctuations of $\Delta \theta$ in such a situation is, up to our
knowledge, unknown and the discovery of the validity of the FR with a
slope similar to a very different regime is largely unanticipated. We
mention that in the Gaussian approximation, i.e. assuming a parabolic
form for the large deviation rate of $\Delta \theta$ or equivalently 
\begin{equation}
f(\Delta \theta) \sim \exp \left[-\frac{(\Delta \theta-\langle \omega \rangle
\Delta t)^2}{2 D \Delta t}\right],
\end{equation}
leads to the identification $s=2\langle \omega \rangle/D$. Again, no
theoretical expectations exist, for the stochastic process modeled in
Eq.~\eqref{beq}, for the ratio between the average drift and the
angular ``diffusion'' coefficient $D$. The empirical observation that
such a ratio is somehow independent from the relative importance
between collisions and dry friction (controlled by $\beta^{-1}$) is
quite an interesting fact. Interestingly, in a previous
paper~\cite{JLM12} where similar measurements have been done on the
different setup cited in the introduction~\cite{EWLM10}, the validity
of the FR for the asymmetry function $\phi(\Delta \theta)$ was
observed at very different shaking strengths, with a slope $\sim 0.2$
independent from the dynamical regime (see Fig. 2a and Fig. 3b of that
paper) and amazingly close to the slope measured in our experiment.

\section{Conclusions}

In summary we have repeated the studies recently appeared
in~\cite{EWLM10} and~\cite{JLM12}, concerning the experimental
measurement of the average Brownian motor effect and the analysis of
the FR respectively, for a new setup which appears to be simple enough
for a reasonable comparison against kinetic theory. Such a theory
predicts two main regimes, rare collisions and frequent collisions,
with two different formula for the average angular velocity of the
rotator, formula~\eqref{drift_rc} and~\eqref{drift_fc}
respectively. The same theory is able, only in the FC, to predict the
validity of the FR $\phi(\Delta \theta)=s \Delta \theta$ with an
analytical formula for the slope $s$, which is in good agreement with
the experiments in that regime. The interesting observation, detailed
above, is that such a slope does not depend strongly upon
$\beta^{-1}$, giving a similar value even in a regime where collisions
are rare excitations followed by fast dissipation due to dry
friction. Future investigation of this puzzling observation is in
order, in particular through numerical simulations or further
variations of the experimental setup.

\begin{acknowledgments}
The authors acknowledge the support of the
Italian MIUR under the grants: FIRB-IDEAS n. RBID08Z9JE, 
FIRBs n. RBFR081IUK and n. RBFR08M3P4, and PRIN  n. 2009PYYZM5.
\end{acknowledgments}

\bibliography{fluct.bib}

\end{document}